\documentclass[sigconf]{acmart}
\usepackage{subcaption}
\usepackage{booktabs}
\usepackage{amsmath}
\usepackage{xcolor}
\usepackage{tikz}
\usetikzlibrary{arrows.meta,positioning}

\let\acmcopyrightfootnote\footnotetextcopyrightpermission
\renewcommand\footnotetextcopyrightpermission[1]{
  \acmcopyrightfootnote{Correspondence to Eun Cheol Choi
  \textless\href{mailto:euncheol@usc.edu}{euncheol@usc.edu}\textgreater.}}

\makeatletter                                         
\newcommand{\preprintnotice}{Preprint. Accepted to ACM Hypertext 2026.}
\AtBeginDocument{%
  \fancypagestyle{standardpagestyle}{%
    \fancyhf{}%
    \renewcommand{\headrulewidth}{\z@}%
    \renewcommand{\footrulewidth}{\z@}%
    \fancyfoot[C]{\if@ACM@printfolios\footnotesize\thepage\fi}%
    \fancyhead[LO]{\@headfootfont\shorttitle}%
    \fancyhead[RE]{\@headfootfont\@shortauthors}%
    \fancyhead[LE]{\@headfootfont\preprintnotice}%
    \fancyhead[RO]{\@headfootfont\preprintnotice}%
  }%
  \pagestyle{standardpagestyle}%
}
\makeatother

\title{Angry but Accurate: Detecting and Profiling the Counter-Misinformation Ecosystem on Twitter}

\author{Eun Cheol Choi}
\orcid{0000-0003-0861-1343}
\affiliation{%
\institution{University of Southern California}
\department{Annenberg School of Communication}
\city{Los Angeles}
\state{CA}
\postcode{90087}
\country{United States}}
\email{euncheol@usc.edu}

\author{Emilio Ferrara}
\orcid{0000-0002-1942-2831}
\affiliation{%
\institution{University of Southern California}
\department{Thomas Lord Department of Computer Science}
\city{Los Angeles}
\state{CA}
\postcode{90087}
\country{United States}}
\email{emiliofe@usc.edu}

\renewcommand{\shortauthors}{Choi \& Ferrara}

\begin{document}

\begin{abstract}
On social media, many users actively push back against false claims. Understanding who pushes back and how they do so matters, as this corrective activity is central to how misinformation is contested. We study this \textit{counter-misinformation ecosystem} at scale: applying a domain-specific NLI model from our prior work to a large corpus of COVID-19 tweets, we classify 264,737 posts as supporting or opposing false claims and compare 23 user- and text-level features across the two groups. Contrary to the dominant assumption that negative emotion is a signature of falsehood, we find that misinformation-opposing posts are \emph{more} emotionally negative than misinformation-supporting posts, with higher levels of anger, disgust, and sadness. These differences are modest in magnitude but consistent in direction across the negative emotions. We also find that posts opposing misinformation tend to come from more established users, i.e., older accounts, more followers, and higher listed counts.
\end{abstract}

\maketitle

\section{Introduction}

Misinformation on social media poses a significant threat to public discourse and informed decision-making, but not all online activity contributes to its spread. Many users actively push back, correcting falsehoods in their feeds \cite{ma2023characterizing}, and most such organic correction comes from ordinary users rather than professional fact-checkers \cite{micallef2020role}. These corrections matter: peer correction can reduce belief in the underlying misinformation \cite{bode2018see}, thus platforms have formalized collective pushback through community-based fact-checking \cite{prollochs2022community}. 

The implication is that combating misinformation requires more than suppressing false content. It also requires understanding the corrective content that emerges in response \cite{bode2018see, prollochs2022community, mosleh2021perverse}. Characterizing the corrective populations---which we term the \textit{counter-misinformation ecosystem}---in both linguistic and account-level terms can provide insight into how misinformation is organically challenged on social media \cite{lee2022prevalence, micallef2020role}.

We address this by first \emph{retrieving} candidate posts for each fact-checked claim, \emph{detecting} which posts support or oppose it at scale, and then \emph{profiling} the populations they form. For detection, we apply a Natural Language Inference (NLI) model described in our previous work \cite{choi2025limited} to a large corpus of COVID-19 tweets \cite{chen2020}, yielding 264,737 model-classified tweets that either support or oppose COVID-19 misinformation. For profiling, we compare 23 user- and text-level features previously linked to misinformation spread, including emotion scores, bot-likelihood scores \cite{ferrara2016rise, chang2021social, ferrara2020characterizing, yang2022botometer}, and toxicity measures \cite{hanley2023sub, pascual2021toxicity, noor2023comparing, quattrociocchi2022reliability}, across the misinformation-supporting and opposing tweets.\footnotemark 

We find that posts \emph{opposing} misinformation are more emotionally charged than those \emph{spreading} it, with higher levels of anger, disgust, and sadness. This runs against the dominant assumption that negative emotion is primarily a feature of false rather than corrective content \cite{weismueller2024information, hosseini2023emotional, solovev2022moral, mcloughlin2021role}; our study corroborates, at scale and with discrete-emotion measurement, prior observations on a few COVID-19 topics \cite{micallef2020role}, alongside a study documenting emotional complexity across both misinformation and corrections during mass-shooting events \cite{lee2022prevalence}. The counter-misinformation ecosystem is also anchored by more established users (older accounts, more followers, and listed more often), and bot scores do not meaningfully differentiate the two sides.

\footnotetext{All data are drawn from 2020--2021, during the COVID-19 pandemic and prior to the platform's rebranding to X; we therefore refer to the platform throughout as Twitter.}

\section{Prior Work}

\subsection{User Corrections of Misinformation}

Beyond how misinformation spreads, a growing literature asks how ordinary users push back against it. Much of this organic correction comes not from professional fact-checkers but from ordinary users: in a study of COVID-19 misinformation on Twitter, 96\% of refuting posts were authored by concerned citizens rather than fact-checking organizations \cite{micallef2020role}. 

Prior RCT work shows that correction by fellow users can be effective in lowering false beliefs \cite{bode2018see}. Its effectiveness may be socially contingent, however: corrections are far more likely to be accepted when they come from someone socially tied to the corrected user than from a stranger \cite{margolin2018political}, and public correction can even backfire downstream, increasing the partisanship and toxicity of a user's subsequent sharing behavior \cite{mosleh2021perverse}. Even so, early fears that repeated correction would broadly entrench false beliefs through a ``backfire effect'' have largely failed to replicate on average \cite{swire2020searching}, and platforms have formalized collective pushback through community-based fact-checking \cite{prollochs2022community}. 

\subsection{Characteristics of Misinformation-Related Content}

Emotional tone is one of the most-studied features distinguishing misinformation from accurate content. A consistent line of work finds that false content is more emotionally charged than truthful content, carrying higher moral outrage, fear, and other affective cues that may encourage sharing \cite{weismueller2024information, hosseini2023emotional, solovev2022moral, mcloughlin2021role}. The picture, however, is not simply ``false = emotional, true = neutral'': accurate information can also provoke strong emotional reactions, and emotional intensity alone is not a reliable truth signal \cite{vosoughi2018spread}. Most directly, a study of COVID-19 fact-check responses on Twitter found that ordinary users' opinion-based rebuttals carried more negative emotion and anger than the misinformation they targeted \cite{micallef2020role}, and a manual study of mass-shooting crises likewise found that correction tweets featured anger \cite{lee2022prevalence}. These observations, however, rested on a limited number of false claims and on dictionary-based or manual coding, leaving open whether the pattern scales and how it appears under discrete-emotion measurement. These limitations leave unanswered whether emotional expression systematically differs between posts that \emph{support} misinformation and posts that \emph{challenge} it.

Beyond emotion, researchers have examined toxic language and automated accounts. Toxic or aggressive posts have been linked to false claims \cite{hanley2023sub, pascual2021toxicity, noor2023comparing}, though the association is inconsistent and not always significant \cite{quattrociocchi2022reliability}. Bot-driven amplification of disinformation and conspiracy content is well documented on Twitter \cite{ferrara2016rise, chang2021social, ferrara2020characterizing, yang2022botometer}. Together, these threads argue for understanding the spread of misinformation in terms of emotion, linguistic style, and account-level characteristics, not content alone.

\subsection{Misinformation Detection}

Misinformation detection (MID) identifies false or misleading content in large online corpora, typically by linking fact-checker resources to social media posts. Common linkage strategies have notable blind spots \cite{choi2024automated}: URL-based sampling misses posts that spread falsehoods through text alone \cite{wirtschafter2024detecting, popat2017truth}, and hashtag-based sampling captures only posts that include specific topic tags, missing the rest \cite{rafail2018, chen2021}.

To address these gaps, researchers have increasingly turned to natural language inference and stance detection, which are used to classify whether a text supports, opposes, or is neutral toward a claim. A range of methods have been applied, from earlier neural architectures to pretrained language models \cite{borges2019combining, dulhanty2019taking}, and incorporating textual entailment features directly into fake-news detection systems can improve classification \cite{sadeghi2022fake, saikh2019novel, yavary2019information}. Open challenges around generalization, domain adaptation, and scalability remain \cite{hardalov2021survey}.

Misinformation generation using LLMs has been proposed as a tool for improving \emph{detection}: synthetic false content, controllably labeled, can expand training sets without the cost of manual annotation \cite{satapara2024fighting, chen2023combating, choi2025limited}. A practical limitation is that commercial LLMs frequently refuse to generate sensitive content (e.g., fake news, profanity), thereby reducing both the volume and diversity of augmentation data \cite{ding2024data, piedboeuf2023chatgpt}. Working with LLM-generated data in this domain, therefore, requires attention to the tension between model safety and research utility.

\section{Methods} \label{sec:methods}

\begin{figure}[h]
\centering
\begin{tikzpicture}[
  font=\normalsize,
  node distance=3.2mm,
  box/.style={rounded corners=2pt, draw, align=left, inner sep=4pt,
              text width=\dimexpr\columnwidth-16pt\relax},
  new/.style={box, fill=cyan!12, draw=cyan!55!black},
  arr/.style={-{Latex[length=2mm,width=1.6mm]}, semithick}
]
\node[new] (c1) {\textbf{1. Retrieval} \\
  Aggregate fact-checked false COVID-19 claims (\S\ref{sec:claims_data}); BM25 retrieves candidate tweets per claim (\S\ref{sec:pairs_data}).};
\node[new, below=of c1] (c2) {\textbf{2. Detection} \\
  Fine-tuned NLI model (\S\ref{sec:finetuning}, \cite{choi2025limited}) labels each pair, resulting in 264,737 tweets supporting or opposing false claims (\S\ref{sec:classification})};
\node[new, below=of c2] (c3) {\textbf{3. Profiling} \\
  Compare user/text features: tweets supporting vs. opposing false claims (\S\ref{sec:profiling})};
\draw[arr] (c1)--(c2);
\draw[arr] (c2)--(c3);
\end{tikzpicture}
\caption{Overview of Methods (\S\ref{sec:methods}).}
\label{fig:pipeline}
\end{figure}

\subsection{Dataset}
\subsubsection{False Claims} \label{sec:claims_data}

We aggregated false COVID-19 claims from several publicly available datasets covering fact-checking platforms, social media, and news outlets, providing broad coverage of the misinformation that circulated during the pandemic (Table~\ref{tab:claim}). After standardizing each source's ``false'' label, filtering for claims marked false or fabricated, and removing duplicates, we obtained 15,374 unique COVID-19 false claims.

\begin{table}[b]
\centering
\caption{False-claim datasets aggregated. Final deduplicated corpus contains 15,374 unique claims. Gf: Google Fact Check Tools; Pf: PolitiFact; Py: Poynter; Sn: Snopes.}
\label{tab:claim}
\small
\begin{tabular}{lllc}
\toprule
\textbf{Source dataset} & \textbf{Origin} & \textbf{Label used} & \textbf{Count} \\
\midrule
FACT-GPT \cite{choi2024fact} & Gf+Pf & --- & 1{,}225 \\
COVID-Fact \cite{saakyan2021covid} & Reddit & REFUTED & 2{,}790 \\
FaCov \cite{sharma2022facov} & Gf & false & 3{,}089 \\
FakeCovid \cite{shahi2020fakecovid} & Py+Sn & FALSE/false & 6{,}249 \\
CoAID \cite{cui2020coaid} & Py+Sn & Claim/NewsFake & 925 \\
FibVID \cite{kim2021fibvid} & Sn+Pf & fake & 569 \\
COVID-19 Rumor \cite{cheng2021covid} & News & F & 3{,}681 \\
COVERT \cite{mohr2022covert} & Twitter & REFUTES & 66 \\
CMU-MisCov19 \cite{memon2020cmu} & Twitter & fake*/false* & 17 \\
\bottomrule
\end{tabular}
\end{table}

\subsubsection{Retrieving Claim--Tweet Pairs} \label{sec:pairs_data}

To link claims to real-world posts, we drew on the large-scale public Coronavirus Twitter dataset \cite{chen2020}, which covers 2020 to 2021. For each of the 15,374 false claims (\S\ref{sec:claims_data}), we used BM25 \cite{robertson2009probabilistic} to retrieve the 100 most relevant tweets, yielding 1,537,400 candidate claim and tweet pairs for downstream classification.

\subsection{Analytical Procedure}

\subsubsection{Classifier Training and Evaluation.} \label{sec:finetuning}
We use the previous fine-tuning dataset and the two-step fine-tuning process described in \cite{choi2025limited}, which we briefly summarize here. The base model was a general-purpose NLI cross-encoder \cite{reimers2019sentence} built on DeBERTa-v3-base \cite{he2021deberta}, pretrained on SNLI \cite{bowman-etal-2015-large}, MNLI \cite{N18-1101}, and FEVER-NLI \cite{nie2019combining}. We then fine-tuned on domain-specific, COVID-19 misinformation NLI datasets \cite{choi2024fact, hossain2020covidlies, hou2022covmis}. To expand our limited labeled set, we used controllable misinformation generation (CMG) \cite{chen2023can, choi2025limited}: given a false claim and a target label (entailment, contradiction, or neutral), an LLM is prompted to produce a realistic tweet matching that label, and the resulting synthetic pairs are appended to the fine-tuning data. The best model, which was fine-tuned on the GPT-4o-augmented dataset on top of general and domain-specific NLI datasets, showed $F1_{\text{Macro}} = .805$ and $Accuracy = .828$, with label-wise performance being $F1_{\text{Ent}} = .779$, $F1_{\text{Con}} = .745$, and $F1_{\text{Neu}} = .885$ on a test set.

\subsubsection{Labeling Tweets.} \label{sec:classification}
We applied the classifier to the full set of 1,537,400 claim-tweet pairs. After removing duplicate tweets, tweets classified as neutral, and filtering ambiguous cases (the same tweet classified as supporting one claim but opposing another), the final analytical sample comprised 264,737 tweets (195,922 supporting and 68,815 opposing false claims) authored by 201,334 unique users.

\begin{figure}[t]
    \centering
    \includegraphics[width=.95\linewidth]{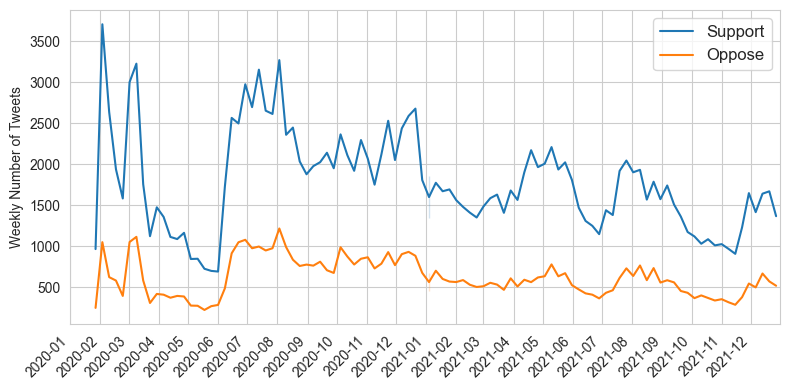}
    \caption{Weekly volume of retrieved tweets supporting or opposing misinformation across 2020 and 2021.}
    \label{fig:weekly_trend}
\end{figure}

\begin{figure*}[t]
    \centering
    \includegraphics[width=.66\linewidth]{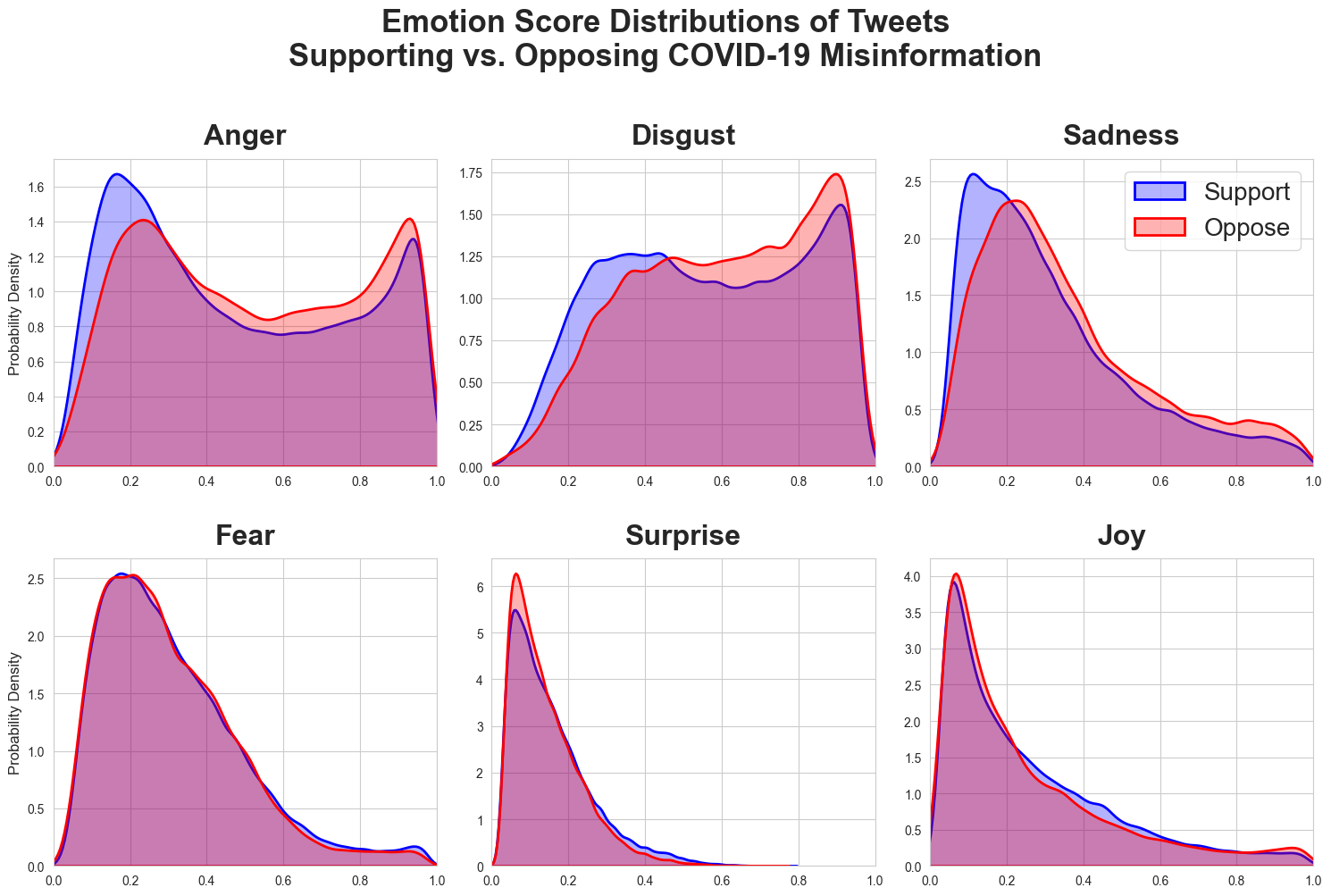}
    \caption{KDE distributions of Ekman's six emotion scores in misinformation-supporting vs. opposing tweets. Opposing tweets exhibit higher levels of anger, disgust, and sadness.}
    \label{fig:teaser}
\end{figure*}

\subsubsection{Profiling Misinformation-Supporting/Opposing Tweets.} \label{sec:profiling}
This post-detection sample lets us move beyond classification to compare \emph{how} misinformation-supporting and opposing tweets are expressed and \emph{who} posts them. We compared 23 user- and text-level features between the two groups: emotional tone via the Demux model \cite{chochlakis2023leveraging}, which scores Ekman's six basic emotions (\textit{anger, disgust, fear, joy, sadness, surprise}) \cite{ekman1992there}, since emotion is among the most-studied correlates of false-information spread \cite{lee2022prevalence}; toxicity via DeToxify \cite{Detoxify}, which produces seven scores (\textit{toxicity, severe toxicity, obscene, identity attack, insult, threat, sexually explicit}); and bot likelihood via the Botometer X API \cite{davis2016botornot, yang2020scalable}, given prior evidence that automated accounts amplify disinformation \cite{ferrara2020characterizing}. Additional user features (\textit{friends, listed count, followers, likes, tweets/retweets, account age}) and text features (\textit{tweet length, hashtag and mention counts}) were drawn directly from the Twitter dataset \cite{chen2020}.

\begin{table}[b]
    \centering
    \caption{User and textual features with $|\delta| > 0.05$. Negative $\delta$ indicates a higher feature value in misinformation-opposing tweets.}
    \label{tab:cliffs_delta_comparison}
    \small
    \begin{tabular}{lcc}
    \toprule
    \textbf{Variable} & \textbf{Type} & \textbf{Cliff's $\delta$} \\
    \midrule
    Sadness & Text & $-0.125$ \\
    Anger & Text & $-0.098$ \\
    Disgust & Text & $-0.097$ \\
    Listed & User & $-0.072$ \\
    Tweet length & Text & $\phantom{-}0.067$ \\
    Followers & User & $-0.058$ \\
    Surprise & Text & $\phantom{-}0.058$ \\
    Account age & User & $-0.052$ \\
    \# of Tweets + RTs & User & $-0.051$ \\
    \bottomrule
    \end{tabular}
\end{table}

\section{Results}

Figure \ref{fig:weekly_trend} presents the weekly distribution of tweets classified as supporting or opposing misinformation over the course of 2020 and 2021. Table~\ref{tab:cliffs_delta_comparison} ranks top features by the absolute value of Cliff's $\delta$ \cite{cliff1993dominance}, a nonparametric measure of effect size, when comparing tweets that support misinformation to those opposing it.

\subsection{Emotional Patterns: The Angry Correctors}
Our central finding concerns the emotional content of tweets supporting/opposing misinformation. Figure \ref{fig:teaser} displays the distributions of Ekman's six basic emotion scores across the two groups. The differences are most pronounced for three negative emotions: sadness, anger, and disgust. In each case, tweets opposing misinformation score higher than those supporting it. This pattern is also illustrated in Table \ref{tab:cliffs_delta_comparison}. Sadness ($\delta$ = -0.125), anger (-0.098), and disgust (-0.097) showed the largest effect. The negative sign indicates that these emotions are more prevalent in counter-misinformation tweets. In absolute terms, these effects are modest ($\lvert\delta\rvert \approx 0.1$), yet they are consistent across all three negative emotions. Conversely, misinformation-supporting tweets scored slightly higher on surprise ($\delta$ = 0.058) and ran somewhat longer ($\delta$ = 0.067).

\begin{figure}[t]
    \centering
    \includegraphics[width=.85\linewidth]{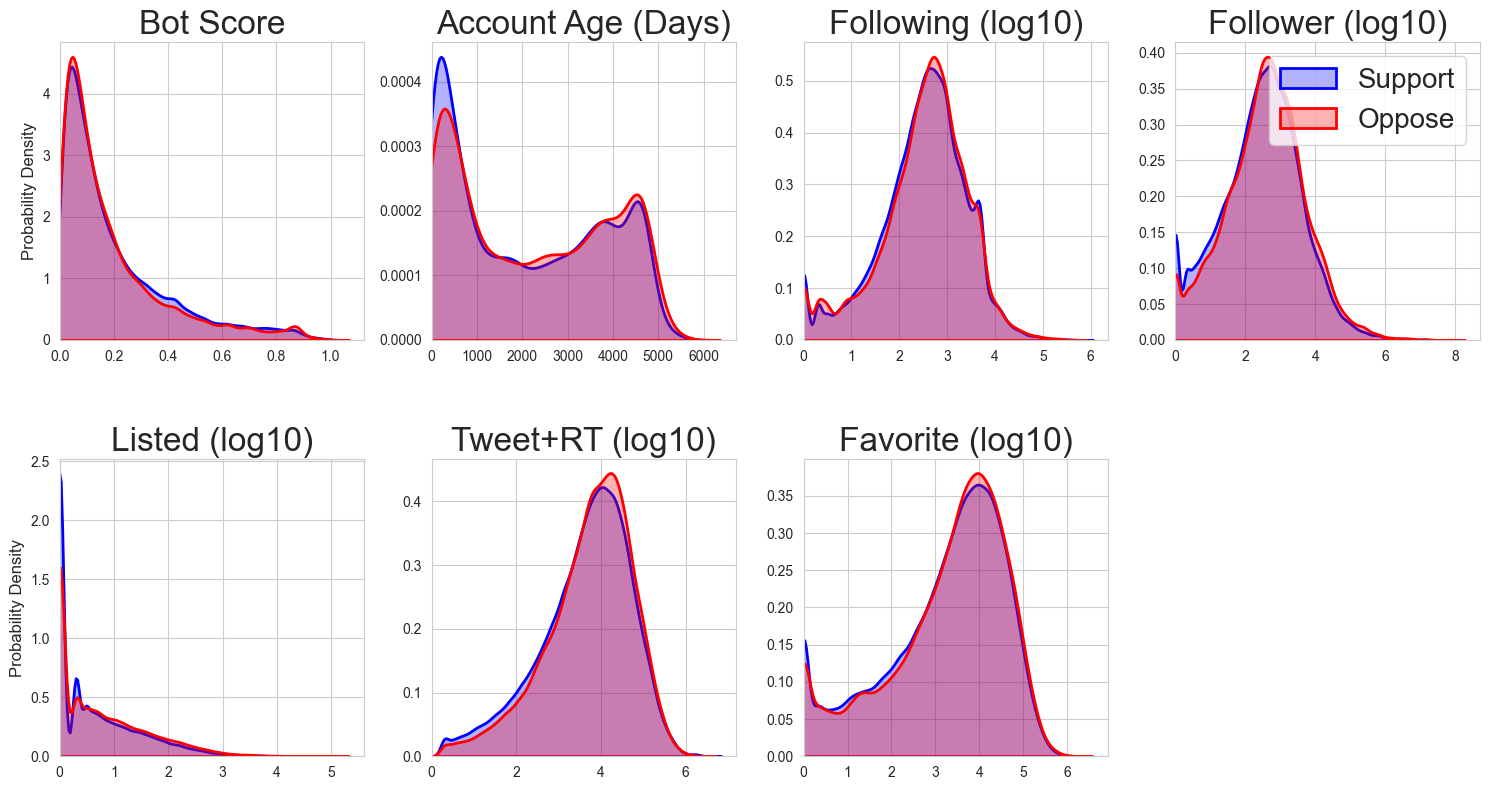}
    \caption{Distributions of user attributes across tweets supporting vs. opposing misinformation.}
    \label{fig:user_attribute}
\end{figure}

\subsection{User Characteristics: Who Pushes Back?}
\begin{figure}[t]
    \centering
    \includegraphics[width=.85\linewidth]{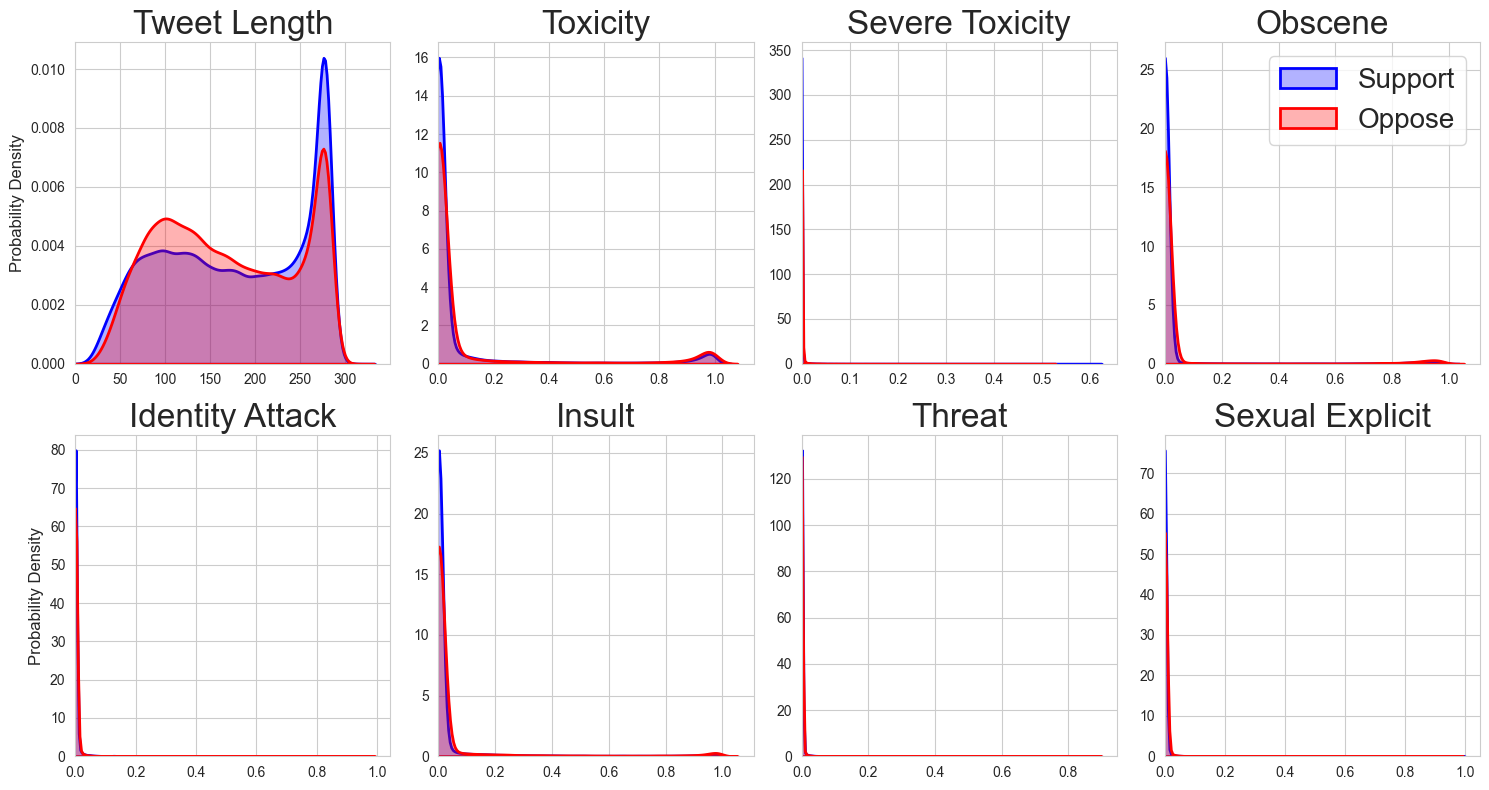}
    \caption{Distributions of text attributes across tweets supporting vs. opposing misinformation.}
    \label{fig:text_attribute}
\end{figure}

Beyond the tweets' content, we also found meaningful differences in who is saying it (Table~\ref{tab:cliffs_delta_comparison}). Users who posted content opposed to misinformation tended to have more followers ($\delta$ = -0.058), were more likely to appear on Twitter lists ($\delta$ = -0.072), had older accounts ($\delta$ = -0.052), and showed higher overall activity levels in terms of tweets and retweets ($\delta$ = -0.051). Notably, bot scores did not emerge as a strongly differentiating feature between the two groups. Distributions for all user and text attributes are reported in Figure~\ref{fig:user_attribute} and Figure~\ref{fig:text_attribute}.

\subsection{Predictive Modeling: Can We Tell Them Apart?}
To move beyond descriptive comparisons, we trained several machine learning classifiers to predict whether a tweet opposes or supports misinformation based solely on the 23 user and textual features described above. Because the two classes are imbalanced (195,922 supporting vs.\ 68,815 opposing), we undersampled the majority class to a balanced set for training and evaluation, so chance accuracy is 0.5. As shown in Table \ref{tab:tree_performance}, random forest achieved the best overall performance, with an F1 score of 0.632 and accuracy of 0.620. XGBoost performed comparably, with slightly higher precision (0.622) but lower recall. Linear models performed noticeably worse, suggesting that the relationship between features and stance is not purely linear. This separability is weak but consistently above chance, indicating that user- and text-level features alone carry a real, if limited, signal for distinguishing the two stances, without access to the tweets' semantic content.

\begin{figure}[t]
    \centering
    \includegraphics[width=.85\linewidth]{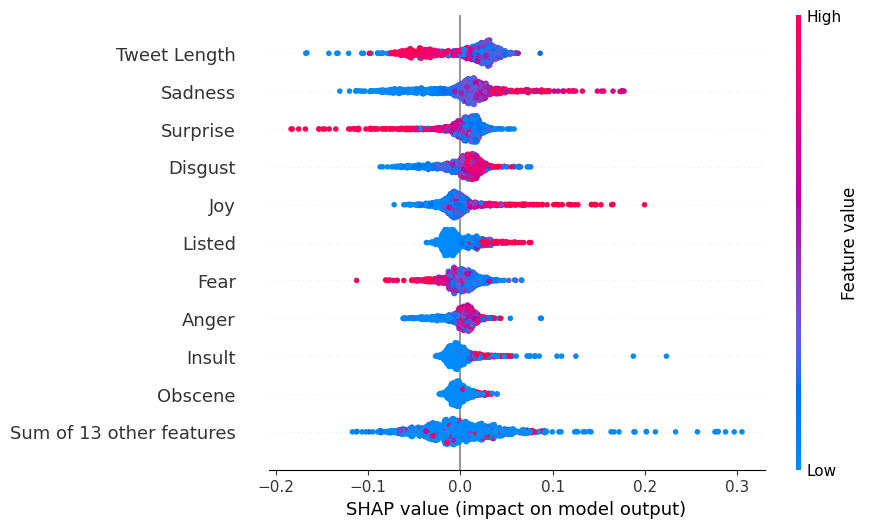}
    \caption{SHAP summary plot showing which features push the random-forest classifier toward predicting \textit{oppose} (positive SHAP) vs.\ \textit{support} (negative SHAP).}
    \label{fig:shap_opposing}
\end{figure}

To interpret the random forest model's decision-making process, we employed SHAP (SHapley Additive exPlanations) values \cite{SHAP}, calculated on a random sample of 1,000 tweets from the dataset. The SHAP summary plot (Figure \ref{fig:shap_opposing}) reveals which features most influenced the model's predictions. Each point in the plot represents a single tweet; its horizontal position indicates how much that feature pushed the model toward predicting ``oppose'' (positive SHAP value) or ``support'' (negative SHAP value). Its color indicates whether the feature value was high (red) or low (blue).

Tweet length emerges as the most influential feature: shorter tweets are more likely to be classified as opposing misinformation. Among the emotional features, sadness and disgust stand out: higher values for these emotions lead the model to predict opposition, corroborating our earlier distributional findings. Surprise works in the opposite direction, with higher surprise scores associated with supporting misinformation. These results reinforce and extend the patterns observed in Cliff's $\delta$ analysis, confirming that the emotional and stylistic signatures of misinformation-supporting and opposing content are distinct enough to be leveraged by machine learning models, even without access to the semantic content of the tweets themselves.

\section{Discussion}
In this work, we characterized the \textit{counter-misinformation ecosystem} by contrasting the posts and users that push back against false claims with those that spread them. Using an NLI-based pipeline fine-tuned on domain-specific COVID-19 data augmented with GPT-4o-generated synthetic pairs, we classified 264,737 tweets as supporting or opposing fact-checked false claims, then compared 23 user- and text-level features across the two groups.

Our most noteworthy finding is that posts \emph{opposing} misinformation are more emotionally charged than those \emph{spreading} it, particularly in sadness, anger, and disgust. This directly challenges the prevailing assumption that negative emotion is primarily a feature of misinformation itself \cite{weismueller2024information, hosseini2023emotional, solovev2022moral, mcloughlin2021role}. Few earlier works had noted that correctors, too, carry negative emotion; a two-topic study found COVID-19 counter-misinformation angrier than the falsehoods it targeted \cite{micallef2020role}, and a manual study of crisis misinformation found correction tweets featured anger \cite{lee2022prevalence}. But the comparative direction, corrective posts being \emph{more} negative than misinformation, had not been established at scale; our contribution is to show it across a broad claim set and with discrete-emotion measurement rather than dictionary-based or manual coding. People who push back against false claims do so with frustration and indignation: they are, in a sense, angry \emph{but} accurate. Accordingly, content-moderation systems that flag emotionally charged posts as likely misinformation may inadvertently suppress the corrective discourse that helps fight it.

\begin{table}[t]
\centering
\caption{Performance of binary classifiers predicting whether a post opposes misinformation, using only user/text features.}
\label{tab:tree_performance}
\small
\begin{tabular}{lcccc}
\toprule
\textbf{Model} & Precision & Recall & $F1$ & Accuracy \\
\midrule
Logistic            & 0.571 & 0.568 & 0.570 & 0.567 \\
Logistic (Lasso)    & 0.571 & 0.568 & 0.569 & 0.566 \\
Logistic (Ridge)    & 0.571 & 0.570 & 0.570 & 0.567 \\
Decision Tree       & 0.556 & 0.551 & 0.553 & 0.551 \\
Random Forest       & 0.618 & \textbf{0.647} & \textbf{0.632} & \textbf{0.620} \\
XGBoost             & \textbf{0.622} & 0.631 & 0.627 & \textbf{0.620} \\
CATBoost            & 0.618 & 0.622 & 0.620 & 0.615 \\
\bottomrule
\end{tabular}
\end{table}

Counter-misinformation users tend to be more established (more followers, older accounts, higher list counts, more activity), consistent with social correction emerging from users with public presence and platform capital; an earlier two-topic study likewise found COVID-19 counter-misinformation accounts older and more followed than misinformation accounts \cite{micallef2020role}, a pattern we confirm across a broader claim set. Bot scores did not separate the two groups, suggesting that automation does not disproportionately concentrate on either side of the divide in our data.

\subsection{Limitations}
Our analysis is restricted to COVID-19 misinformation on Twitter during 2020 and 2021, which may limit generalizability to other topics, platforms, or time periods; the emotional dynamics we observe could partly reflect that period's heightened anxiety and political polarization. Our feature comparison relies on pretrained models (Demux for emotion, DeToxify for toxicity, Botometer for bot detection) that carry their own biases and measurement error. The NLI classifier is also imperfect, and residual misclassification may add noise to the comparative findings. Because we replicate the fine-tuning and CMG augmentation pipeline \cite{choi2025limited}, we additionally inherit two limitations they document: LLM refusal rates (even the best-performing GPT-4o refuses a small fraction of generation requests, leaving certain content types systematically under-represented in the augmentation set) and \emph{task flipping} (LLMs occasionally generate content opposite to the requested stance, introducing noise into training labels), and gains over simpler augmentation baselines remain modest. The group-level contrasts we report are only modest at the level of individual tweets: trained on these features alone, classifiers separate the two stances weakly (Table~\ref{tab:tree_performance}), so the differences reflect distributional tendencies rather than reliable per-tweet signals. We also do not control for user-level demographic or partisanship covariates, which our data do not directly observe; the ``more established users'' pattern on the counter-misinformation side could therefore partially reflect demographic or political-leaning differences between the two populations rather than corrective behavior per se. Finally, our dataset includes only publicly accessible tweets, excluding deleted, private, or suspended-account content that may systematically differ.

\subsection{Implications and Future Research}
For platform governance, the most direct takeaway is that emotional tone is a poor proxy for misinformation: flagging or downranking content based on negative-emotion signals risks silencing legitimate corrective discourse and, with it, an established core of users who anchor the counter-misinformation ecosystem. For the misinformation research community, our work demonstrates the value of NLI-based detection paired with population-level profiling for studying the response side at scale. Future work should extend this analysis to political and climate-related falsehoods to test whether the emotional pattern generalizes; investigate \emph{how} and \emph{why} misinformation-opposing posts are more emotionally charged (genuine frustration, strategic framing, or audience effects); and use longitudinal designs to trace how the misinformation and counter-misinformation populations co-evolve across an unfolding event. Methodologically, more sophisticated integration of NLI and SD, and effective LLM-driven augmentation strategies for sensitive domains, remain promising directions.

\textbf{AI Tools Usage Disclosure}. During the preparation of this work, the authors used ChatGPT and Claude to refine text and review code. The authors reviewed and edited the content as needed and take full responsibility for the publication's content.

\bibliographystyle{ACM-Reference-Format}
\bibliography{bibliography}

\end{document}